\documentclass[12pt]{iopart}

\usepackage{multirow}
\usepackage{graphicx}
\usepackage{iopams} 
 
\begin{document}

\title[tDES for amplitude fluctuations]{Threshold distribution of equal states for quantitative amplitude fluctuations}

\author{Wenpo Yao$^{1,2}$ Wenli Yao$^3$ and Jun Wang$^1$}

\address{$^1$ School of Geographic and Biologic Information, Smart Health Big Data Analysis and Location Services Engineering Lab of Jiangsu Province, Nanjing University of Posts and Telecommunications, Nanjing 210023, China}

\address{$^2$ Key Laboratory of Computational Neuroscience and Brain-Inspired Intelligence (Fudan University), Ministry of Education}

\address{$^3$ State Key Laboratory of Hydroscience and Engineering, Tsinghua University, Beijing 100084, China}

\ead{yaowpo@hotmail.com}

\begin{abstract}
\textit{Objective}. The distribution of equal states (DES) quantifies amplitude fluctuations in biomedical signals. However, under certain conditions, such as a high resolution of data collection or special signal processing techniques, equal states may be very rare, whereupon the DES fails to measure the amplitude fluctuations. \textit{Approach}. To address this problem, we develop a novel threshold DES (tDES) that measures the distribution of differential states within a threshold. To evaluate the proposed tDES, we first analyze five sets of synthetic signals generated in different frequency bands. We then analyze sleep electroencephalography (EEG) datasets taken from the public PhysioNet. \textit{Main results}. Synthetic signals and detrend-filtered sleep EEGs have no neighboring equal values; however, tDES can effectively measure the amplitude fluctuations within these data. The tDES of EEG data increases significantly as the sleep stage increases, even with datasets covering very short periods, indicating decreased amplitude fluctuations in sleep EEGs. Generally speaking, the presence of more low-frequency components in a physiological series reflects smaller amplitude fluctuations and larger DES. \textit{Significance}.The tDES provides a reliable computing method for quantifying amplitude fluctuations, exhibiting the characteristics of conceptual simplicity and computational robustness. Our findings broaden the application of quantitative amplitude fluctuations and contribute to the classification of sleep stages based on EEG data.
\end{abstract}

\noindent{Distribution of equal states; Amplitude fluctuation; EEG; Sleep stage}

\section{Introduction}
Physiological time series convey important information to identify body conditions, and the quantitative assessment of these time series plays an important role in biomedical signal processing. For example, brain activity exhibits pronounced physiological changes under different sleep conditions. Various methods have been proposed to characterize sleep electroencephalography (EEG) from the perspectives of time and frequency domains, nonlinear dynamics, and coupling causality \cite{Motame2014,Liu2016,Zhao2019}. 

A manual of standardized criteria has long been proposed by Rechtschaffen and Kales (R\&K) \cite{Rechtsch1968} based on EEG and related physiological signals, and standardized scoring manuals are updated annually by American Academy of Sleep Medicine (AASM) \cite{AASM} using sleep polygraphic records. Time- and frequency-domain parameters \cite{Fraiwa2012,Bajaj2013,Miskov2019} have long been used for the classification of sleep stages. Given the inherent complexity of EEG data, nonlinear approaches \cite{Zhao2019} such as entropy measures \cite{Bandt2017}, fractal dimensions \cite{Miskov2019,Ma2018}, and coupling causality \cite{Faes2015} have been widely applied to determine sleep conditions. Currently, machine learning techniques based on the random forest classifier \cite{Fraiwa2012}, convolutional neural networks \cite{Khali2021}, and support vector machines \cite{Liu2016,Lajnef2015} are being introduced to classify sleep EEG data. These methods detect signal features from different perspectives to characterize brain conditions while sleeping. Among these signal features, the most intuitive characteristic of sleep EEG data is the fluctuating amplitude. To quantify amplitude fluctuations, an extremely simple parameter called the distribution of equal states (DES) \cite{Yao2021DES} has been developed in consideration of the limitations of analog-to-digital conversion (ADC).

Coarse-grained ADC can result in different signals falling into the same quantization interval and being recorded as the same value during digital signal collectio. Based on the probability of neighboring equal values, the DES effectively characterizes physiological or pathological information, e.g., in epileptic brain data and cardiac signals \cite{Yao2021DES,Yao2019ER}. The decreased DES of epileptic ictal EEG data is in line with the abnormally large amplitude fluctuations of neuronal firing, and the high DES of postictal EEG data significantly distinguishes the postictal condition from other brain states \cite{Yao2021DES,Cui2019}. Unlike EEG data, heart rate data are generally represented by R waves, which are indirectly derived from electrocardiography, and the increased DES of heartbeats could serve as an independent indicator for potential cardiac mortality \cite{Yao2019ER}. However, as it relies solely on equal states, DES fails to measure biomedical amplitude fluctuations when equal states are very rare. Equal digital data imply zero-amplitude fluctuations, the direct cause of which lies in the coarse-grained quantization process of ADC \cite{Walden1999,Proakis2006,Gray1998}. At higher resolutions, there might be very rare or even no equal data in a signal recording, invalidating the DES. Some signal processing methods (e.g., filtering, detrending, ﻿blind source separation, and denoising) can eliminate the equal states. Additionally, in some information theories, signals are assumed to have a continuous distribution \cite{Cover2006,Rojo2018}. In this case, the rare equal states in a series can be removed by adding small random perturbations or arranging the data in order of occurrence \cite{Bandt2002}. Therefore, the original DES, which relies on the occurrence of equal states in signals, cannot reliably extract biomedical information on amplitude fluctuations in situations where equal states are rare.

To address these limitations of the original DES, we develop a novel threshold DES (tDES) of quantitative amplitude fluctuations in which low-pass filtering is applied to differential states. In this paper, we first introduce the background and limitations of the original DES, and then describe how equal values are transformed into zero differential values to derive the improved tDES. Synthetic signals and real-world sleep EEG data are used to evaluate the new tDES. Based on the results, we discuss symbolic DES, nonlinear dynamics and the limitations of tDES. Overall, our research makes the following contributions: (1) the proposed tDES extends the original DES to have a broader practical significance in biomedical quantitative amplitude fluctuations; and (2) our research improves the reliability of DES and contributes to an improved understanding of sleep stage classification.

\section{Methods}
This section introduces the original DES and describes its drawbacks, before presenting solutions for the quantitative identification of amplitude fluctuations.

\subsection{Original DES and its drawbacks}
\begin{figure}[htb]
	\centering
	\includegraphics[width=8cm,height=6.5cm]{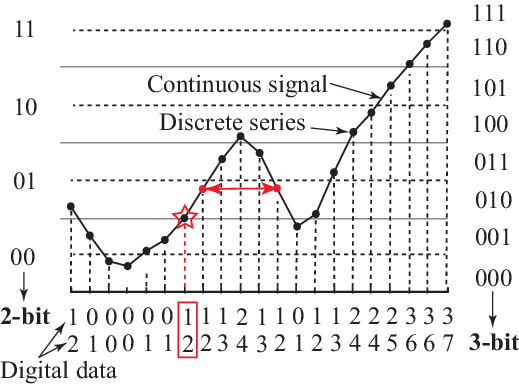}
	\caption{Illustration of an example ADC procedure. A continuous signal is converted into a discrete series (black sampling points) and then mapped into 2-bit (four quantization intervals, on the left) and 3-bit (eight quantization intervals, on the right) digital data. A rounding method is applied that quantizes a value as the nearest digital number. The midpoint element marked by the star is transferred to the larger values of 1 and 2 in the case of 2- and 3-bit digital data, respectively. The red double-arrow line links a pair of equal discrete data.}
	\label{fig1}
\end{figure}

Real-world signals, such as those resulting from physiological activities, speech, and meteorological phenomena, are generally continuous and can be transformed into digital data for computational processing through ADC \cite{Yao2021DES,Walden1999,Proakis2006,Gray1998}. The process of ADC includes sampling, quantization, and coding, among which quantization is nonlinear and irreversible. During the coarse-grained quantization process, elements within same quantitative interval are converted into same digital values, resulting in the quantization error \cite{Walden1999,Proakis2006,Gray1998,Yao2021DES}. Figure~\ref{fig1} illustrates an ADC process with 2- and 3-bit resolution.

In Fig.~\ref{fig1}, the discrete series has one pair of equal values, while the 2-bit digital data contain 12 pairs of neighboring double-equal values and the 3-bit data have 5 pairs. The higher the resolution of ADC, the smaller the quantization error and less equal values. Amplitude is the most direct state of the time series subject to various physiological and pathological conditions. The `state' in DES could be the direct digital value of the series or some indirect features derived from the series, such as the space vector \cite{Yao2021CSF,Kim1999}, order pattern \cite{Bandt2017,Bandt2002,Yao2022PLA,Yao2023}, or partition-based symbol \cite{Daw2003,Lin2007}. In this paper, we focus on the direct digital amplitudes to characterize physiological series. Given a process $X(i)=\{s_{1},s_{2},\ldots,s_{i},\ldots,s_{L}\}$ of length $L$, we target the double-equal states $s_i=s_{i+\tau}$ for computational reasons and natural choices, and we define the DES \cite{Yao2021DES} using Eq.~(\ref{eq1}), where $N(s_i=s_{i+\tau})$ is the number of neighboring equal states and $\tau$ is the time delay.

\begin{eqnarray}
	\label{eq1}
	DES=\frac{N(s_i=s_{i+\tau})}{L-\tau}
\end{eqnarray}

Under an acceptable signal collection resolution, a series with larger amplitude fluctuations has fewer equal digital values; in contrast, a series with small amplitude fluctuations has more neighboring equalities. Therefore, DES is an effective and simple parameter to characterize the amplitude fluctuations \cite{Yao2021DES,Yao2019ER}. However, if the signal recording resolution is very high or the digital data undergo special processing techniques, equal values might be very rare or may not exist. Under these conditions, the original DES fails to extract reliable information about amplitude fluctuations.

\subsection{Threshold DES}
Given the problem that arises when equal states are very rare, we target the distribution of differential values within a given threshold to measure the amplitude fluctuations. This concept of tDES is the central contribution of this paper.

For a process $X(i)=\{s_{1},s_{2},\ldots,s_{i},\ldots,s_{L}\}$, the differential series is $X(i)^{'}=\{s_{1}^{'},s_{2}^{'},\ldots,s_{i}^{'},\ldots,s_{L-\tau}^{'}\}$, where $s_{i}^{'}=s_{i+\tau}-s_{i}$. The tDES is calculated using Eq.~(\ref{eq2}), where $L$ is the length of $X(i)$, $\tau$ is the delay, and $r$ is the threshold.

\begin{eqnarray}
	\label{eq2}
	tDES=\frac{N(|s_{i}^{'}| \leq r)}{L-\tau}
\end{eqnarray}

The original DES is a special case of tDES with $r$=0. The threshold for tDES can be a fixed value or an adaptive value extracted from the signal. For example, in heartbeat analysis, $r$=50 ms is a commonly accepted threshold for the time domain parameter \cite{Yao2018Md,Shaffer2017,Malik1996}. To set an adaptive threshold for tDES, the tolerance applied for information-theoretic probability estimation \cite{Yao2021CSF,Xiong2017} is a preferable choice, particularly for a complex process with unknown characteristics. The tolerance changes according to the standard deviation of the process, as given in Eq.~(\ref{eq3}), where $\bar{s}$ and $L$ are the mean and length of $X$ respectively, and $\alpha$ is an adjustable parameter.

\begin{eqnarray}
	\label{eq3}
	r=\alpha*\sqrt[2]{\sum{(s_{i}^{'}-\bar{s})}/L}
\end{eqnarray}

As it targets the distribution of zero differences, the original DES is an extreme methodology that essentially becomes invalid in the case of very rare equal states. Measuring the distribution of differential states that are no larger than a threshold is a more reasonable and applicable strategy.

\section{Results}
In this section, we generate synthetic signals to evaluate the proposed tDES. The tDES methodology is then applied to the analysis of real-world sleep EEG recordings.

\subsection{Threshold DES in synthetic signal}
Let us first evaluate the proposed tDES using synthetic signals containing different frequency components. Based on groups of sine functions, we generate five sets of synthetic signals under $\delta$ (0.3--3.5 Hz), $\theta$ (4--7 Hz), $\alpha$ (8--13 Hz), $\beta$ (14--30 Hz), and $\gamma$ (31--50 Hz) frequency bands. Because these synthetic data have no neighboring equal values, we apply tDES to quantify their amplitude fluctuations. Synthetic data are normalized (i.e., standard deviations are all 1), so $r$=$\alpha$ in Eq.~(\ref{eq3}). The tDES results for synthetic signals with $r$=$\alpha$=0.01 are listed in Table~\ref{tab1}.

\begin{table}[htb]
	\centering
	\caption{The tDES of 5 sets of synthetic signals}
	\label{tab1}
	\begin{tabular}{cccccc}
		\hline
		\multirow{2}{*}{tDES} &$\delta$	&$\theta$	 &$\alpha$ &$\beta$ &$\gamma$\\
		 &0.3-3.5 Hz	&4-7 Hz	 &8-13 Hz &14-30 Hz	 &31-50 Hz\\
		\hline
		$\tau$=1   &0.2960	&0.1620	&0.0892	&0.0320 &0.0204 \\
		$\tau$=2   &0.1440	&0.0792	&0.0436	&0.0212 &0.0156 \\
		$\tau$=3   &0.0948	&0.0528	&0.0312	&0.0084 &0.0068 \\
		\hline
	\end{tabular}
\end{table}

As frequency band increases, tDES values of the five sets of synthetic signals decrease. Synthetic signals  that contain more high-frequency components exhibit larger amplitude fluctuations, thus giving a smaller tDES. Mathematically, the slope or gradient of a series directly reflects the instantaneous change in amplitude. For digital signals, differential data (i.e., the derivative) correspond to the slope or gradient. The frequency components of a physiological series determine its differential series. Therefore, by measuring the frequency components, tDES could quantify the amplitude fluctuations in physiological series. According to table 1, time delays affect the tDES of a time series. As delay factor increases, the length of neighboring samples increases, i.e., the number of synthetic data samples decreases; therefore, tDES values of the synthetic signals decrease.

\subsection{Sleep EEGs and the original DES}
Sleep EEGs are digital data converted from ionic current voltage fluctuations within brain neurons, which exhibit pronounced changes under different sleep conditions. In this paper, we employ sleep EEGs from the public MIT-BIH polysomnographic database \cite{Ichima1999,Goldber2000} to further test the proposed tDES. These sleep physiological signals were obtained from 16 male subjects aged 32--56 (mean age 43), with weights ranging from 89--152 kg (mean weight 119 kg). In the sleep physiological series, EEG data were recorded from the C4-A1, O2-A1, or C3-O1 channel with a sampling frequency of 250 Hz and 12-bit quantization. The database contains over 80 h of EEG data annotated with respect to sleep stages according to the criteria of Rechtschaffen and Kales \cite{Rechtsch1968} at 30-s intervals. According to expert annotations relating to sleep stages, we extracted sleep data from four stages, i.e., awake and sleep stages 1 (S1), 2 (S2), and 3 (S3). For each stage, we selected 50 sets of EEG segments and cut out 15000 points (60 s) after visual inspection for artifacts. Detailed information about the polysomnographic database can be found in \cite{Ichima1999,PhyNet}. To test the statistical differences in DES values of these EEGs, we conducted nonparametric Mann–Whitney U tests on the DES of each pair of sleeping EEGs and used the Kruskal–Wallis analysis of variance test to measure the statistical differences in the tDES values for the four groups of EEGs. 

Examples of raw sleep EEGs and DES are illustrated in Fig.~\ref{fig1}(a) and \ref{fig1}(b). Next, the EEG data were detrended and filtered through a 0.3–35 Hz bandpass filter using a Butterworth filter in Matlab. The detrend-filtered sleep EEGs are displayed in Fig.~\ref{fig1}(c); note that they have no neighboring equal values. Furthermore, each individual detrending and filtering process eliminates equal values in sleep EEGs.

\begin{figure}[htb]
	\centering
	\includegraphics[width=16cm,height=5cm]{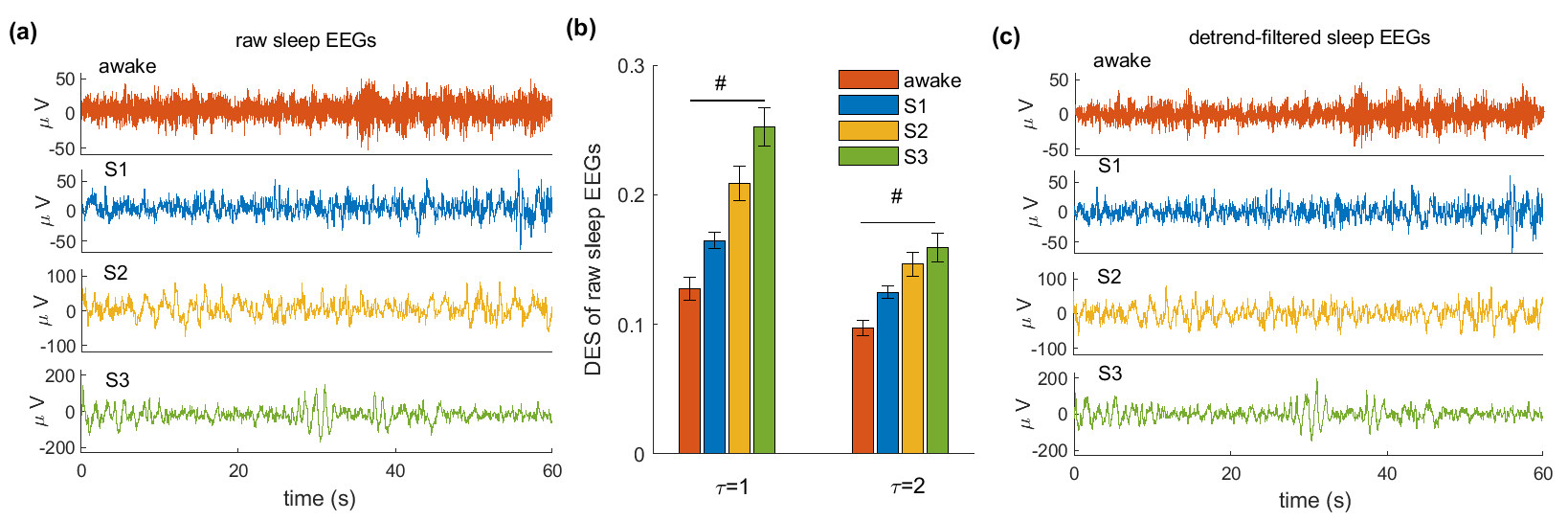}
	\caption{Exemplary raw and detrend-filtered EEG time series and the DES of sleep EEGs (mean$\pm$standard error). The DES of 50 groups of raw sleep EEGs in the awake, S1, S2, and S3 stages when $\tau$=1 and 2 are shown in (b), while those of the detrend-filtered sleep EEGs are all zeros. \# indicates p$<$0.0001 across all stages using the Kruskal--Wallis test. The p-values of the Mann--Whitney test between each pair of sleep EEGs are all smaller than 0.05, except that of S2 and S3 (p=0.386) when $\tau$=2.}
	\label{fig2}
\end{figure}

The raw sleep EEGs contain a large number of neighboring equal values, and DES of these EEG data increases significantly as the sleep stage changes from awake to S1, S2, and S3. The 12-bit quantization in sleep EEGs has much higher resolution than the 2- and 3-bit ADC in Fig.~\ref{fig1}; however, it is still a coarse graining process and discrete series with small amplitude fluctuations are transformed into equal values. In Fig.~\ref{fig2}(a) and \ref{fig2}(c), the awake EEGs contain dramatic wave variations. As the subject begins to sleep and as the sleep depth increases, the DES increases with the decrease of amplitude fluctuations in sleep EEGs. In Fig.~\ref{fig2}(b), the raw sleep EEGs generally contain equal values, e.g., when $\tau$=1, the awake EEGs have about 12\% DES and the S3 EEGs contain approximately 25\% neighboring equal values.

According to Fig.~\ref{fig2}(b), DES values of the sleep EEGs decrease with increasing delay $\tau$, similar to the results using synthetic signals in Table~\ref{tab1}. An increase in the delay is equivalent to (although not identical to) decreasing the signal sampling frequency, in that both reduce the number of sampling points, resulting in a decrease in signal resolution and DES \cite{Yao2021DES}. The time delay should not be too large; otherwise the Nyquist sampling rate (i.e., more than twice the signal band) may not be satisfied or causal disconnection may occur (i.e., irrelevance in the phase space \cite{Kim1999}). At a sampling frequency of 250 Hz, setting $\tau$ to 3 is approximately equivalent to reducing the sampling frequency to 83.3 Hz, resulting in a failure to collect high-frequency brain information. Moreover, the effect of delay on the DES of ADC-direct recordings is complex. It has previously been assumed that the DES value decreases linearly with the delay \cite{Yao2021DES}; however, this is not the case. For example, in the extreme case where the values in a biomedical signal are all the same, DES will remain constant regardless of the delay. Consequently, as signal amplitude fluctuations increase, there will be fewer equal values, and the effect of delay will be greater; on the contrary, if the amplitude variations decrease, the effect of delay will be small. Overall, the effect of delay on DES is related to the sampling frequency of ADC and the frequency components in signals.

\subsection{Threshold DES in sleep EEGs}
\begin{figure}[htb]
	\centering
	\includegraphics[width=16cm,height=11cm]{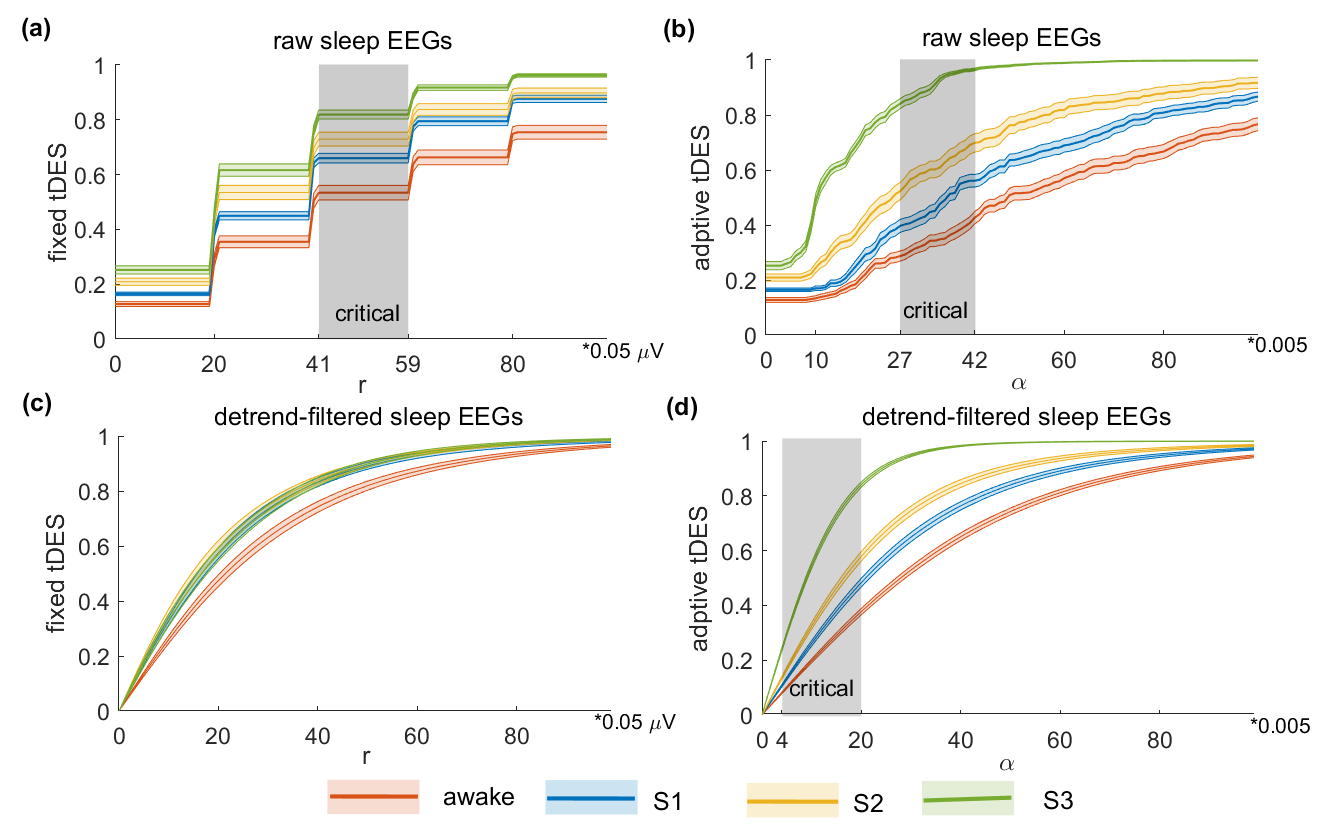}
	\caption{The tDES (mean$\pm$standard error) of the raw and detrend-filtered sleep EEG data. The shaded part in each panel is the critical parameter interval with the optimal statistical differentiation. For the critical $r$ in (a) and $\alpha$ in (b) and (d), the Kruskal–Wallis test results are p$<$0.0001, respectively. The p-values of the Mann–Whitney U test for each pair of sleeping EEG data are all less than 0.005 in (a), (b), and (d).}
	\label{fig3}
\end{figure}

Note that the detrend-filtered sleep EEG data have no equal values. We applied the tDES method to both the raw and detrend-filtered sleep EEG data with increasing fixed and adaptive thresholds. The results are compared in Fig.~\ref{fig3}.

In Fig.~\ref{fig3}(a) and \ref{fig3}(b), tDES of the raw EEGs is initially nonzero, the same as the raw DES, while in Fig.~\ref{fig3}(c) and \ref{fig3}(d), the tDES of the processed EEGs start from zero, that is, there are no equal elements in the detrend-filtered EEGs. The tDES solves the problem of rare equalities in detrend-filtered sleep EEGs. Consistent with the DES of raw EEGs in Fig.~\ref{fig2}, awake EEGs have the smallest tDES, indicating the greatest amplitude fluctuations. The tDES values of EEGs then increase significantly through the sleep stages.

According to statistical results, the adaptive tDES with an adaptive threshold achieves better discrimination with both the raw and detrend-filtered EEG data. For the raw EEG data, the differences between the fixed tDES values of the EEG are most significant (p$<$0.0001) when $r$ is selected from [41, 59]*0.05 $\mu$V in Fig.~\ref{fig3}(a). For adaptive tDES, $\alpha$=36*0.005=0.18 among the interval [27, 42]*0.005 in Fig.~\ref{fig3}(b) is the optimal choice for the tDES. For the detrend-filtered EEG data in Fig.~\ref{fig3}(d), $\alpha$ should be in the interval [4, 20]*0.005 for adaptive tDES to achieve satisfactory discrimination (p$<$0.0001). The Mann–Whitney U test suggests that the tDES with critical thresholds between each pair of sleep EEG data in Fig.~\ref{fig3}(a), \ref{fig3}(b), and \ref{fig3}(d) are all significantly different (p$<$0.005).

The tDES with a fixed threshold presents different outcomes, particularly with the raw EEGs. In Fig.~\ref{fig3}(a), tDES of the raw sleep EEG data increases in steps rather than continuously, and in the initial part of Fig.~\ref{fig3}(b), the tDES of the raw EEG data does not increase because there are only zero differences. We define this phenomenon of step change as amplitude difference level (ADL), and assume the underlying reason might lie in the quantization limitation. Raw signal is transformed into discrete digital values during quantization, and so the differences between the digital values are also distributed discretely. The implication of ADL is that tDES is insensitive to the choice of threshold in the ADL interval because, within this range, there is no change in the tDES of sleep EEG data and their statistical differences.

According to Fig.~\ref{fig3}, tDES effectively distinguishes the raw and detrend-filtered EEG data in different sleep stages, and the adaptive tDES is evidently less sensitive to noise pollution. If the effect of noise is sufficiently small that the amplitude fluctuations remain within the threshold, the value is still unified to 1, whereas if the variations become greater than the threshold, the value is discarded. This coarse-grained processing reduces the effect of noise on quantitative amplitude variations. The coarse-grained processing means that tDES, particularly the adaptive form, is insensitive to noise pollution when measuring the amplitude fluctuations of sleep EEG data.

\begin{figure}[htb]
	\centering
	\includegraphics[width=16cm,height=11cm]{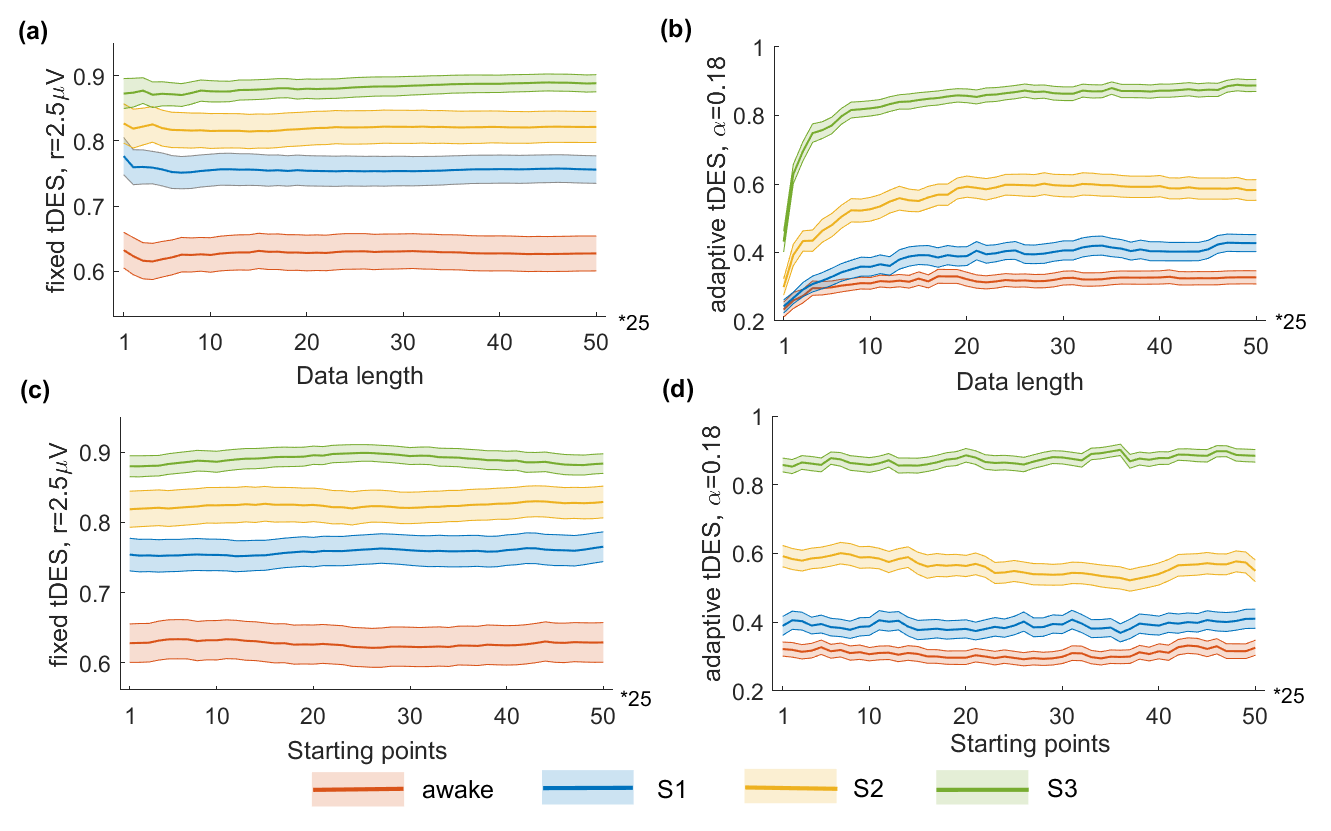}
	\caption{The tDES (mean$\pm$standard error) in the raw sleep EEG data. $r$ of the fixed threshold is 2.5 $\mu$V in (a) and (c), and $\alpha$ of the adaptive threshold is 0.18 in (b) and (d). In (a) and (b), the data length increases in step sizes of 25 (0.1 s). In (c) and (d), the data length is 500 (2 s) and the starting point increases from the initial point in step sizes of 25 points.}
	\label{fig4}
\end{figure}

According to the tDES of the sleep EEG data in Fig.~\ref{fig3}, we select the threshold $r$=50*0.05=2.5 $\mu$V and the control parameter $\alpha$=36*0.005=0.18 in the critical intervals, and we employ the raw sleep EEGs to further test the fixed and adaptive tDES. The tDES performance is shown with respect to the data length and starting point in Fig.~\ref{fig4}.

According to Fig.~\ref{fig4}(a) and \ref{fig4}(b), the tDES, particularly with the fixed threshold, has has very low requirement to data length. Fixed tDES effectively discriminates the four groups of sleep EEG data and even exhibits stationarity after the initial 25 points (Kruskal–Wallis test, p$<$0.0001; Mann–Whitney U test, p$<$0.001). With the adaptive threshold tDES, the four EEGs are significantly different when they contain more than 100 points (Kruskal–Wallis test, p$<$0.0001; Mann–Whitney U test, p$<$0.05). Because the tDES calculation for a time series requires only two neighboring values and very short data lengths can produce stationary quantitative amplitude fluctuations, tDES enables real-time characterization of real-world signals.

From Fig.~\ref{fig4}(c) and \ref{fig4}(d), tDES is insensitive to the starting point of sleep EEG data. With fixed threshold, the tDES of the sleep EEG data is not affected by the initial time point and exhibits high stability. The adaptive tDES shows some overall fluctuations, but the statistical results indicate that effective distinctions are achieved among sleep brain activities.

According to test results, the threshold is a critical parameter in tDES, and should be selected accordingly. If the threshold is too small, the tDES value does not contain much information about the amplitude fluctuations. If the threshold is large, crucial information about the amplitude variations might be overlooked. The adaptive threshold (i.e., the tolerance) is widely employed in information-theoretic approaches \cite{Yao2021CSF,Xiong2017}. Therefore, an adaptive parameter is more suitable for detecting unknown features. The stepped tDES of sleep EEG data confirms the advantages of the fixed threshold, as tDES is insensitive to the selection of the ADL threshold. Therefore, a fixed threshold is more appropriate for some well-known signals, allowing more efficient and simple outcomes; otherwise, the adaptive threshold is the preferable choice.

Overall, tDES based on either a fixed or adaptive threshold effectively addresses the problem of rare equal states in time series. The tDES method inherits the merits of the original DES, i.e., simplicity, low sensitivity to the data length and insensitivity to the starting point.

\section{Discussion}
In this paper, we have proposed the tDES and verified its effectiveness. However, we have found some issues that require further discussion.

Regarding the rare occurrence of equal states, symbolic time series analysis provides another solution. Partition-based symbolization \cite{Daw2003,Lin2007} is a coarse-grained process that exactly resembles ADC, although with different purposes. By dividing the series into several intervals and labeling each segment with a letter, the distribution of equal symbols could be calculated instead of the original values \cite{Yao2021DES}. However, partition symbolization requires assumptions about the size of intervals and segment lines, which increase the complexity of identifying quantitative amplitude variations by symbolic DES. Moreover, partition symbolic DES has the drawback that there might be neighboring elements around the segment lines with differences smaller than the interval, and these will be transformed into different symbols, a problem could be defined as boundary error. With this inherent defect in symbolization, partition-based symbolic DES yields additional errors. Of course, there are other symbolic transformations based on the comparison of neighboring elements, e.g., ordinal patterns \cite{Bandt2017,Bandt2002,Yao2022PLA}, and these are effective in increasing equal states for processes. Overall, by combining symbolic dynamics and quantitative amplitude fluctuations, symbolic DES requires further research as an alternative to the tDES for improving the original DES.

The relationship between amplitude fluctuations and nonlinear dynamics also requires further investigation. In this paper, we have proposed the time-domain tDES of EEGs for sleep classification and made a tentative exploration of its physiological meanings from the frequency domain. In fact, amplitude fluctuations are influenced not only by frequency information, but also by other information such as nonlinear dynamics. The awake EEGs have the highest amplitude fluctuations, and they also have the greatest informational complexity (Shannon entropy is 1.580) among the distributions of differential EEGs; the S1, S2, and S3 EEGs have lower entropy complexities of 1.449, 1.251, and 0.980, respectively. This seems to indicate that increased amplitude fluctuations are correlated with increased complexity, an assumption that is shared by previous analysis \cite{Yao2021DES,Yao2019ER,Yao2018Md}. Meditative heartbeats have smaller temporal fluctuations and joint entropy complexities than non-meditative ones \cite{Yao2018Md}. Healthy young heart rates exhibit high complexity and large amplitude variations, whereas elderly heartbeats, especially from diseased patients, exhibit decreased amplitude fluctuations and less complexity \cite{Yao2019ER,Goldbe2002}. Regarding brain electric activities, epileptic ictal EEGs have abnormally large amplitude fluctuations and nonlinear dynamics \cite{Yao2021DES}. However, postictal brain activity can also manifest very smooth amplitude waves (i.e., high DES) while displaying highly nonlinear dynamics \cite{Yao2021DES,Yao2020ND}. It remains unknown whether postictal brain activities having inconsistent nonlinearity and amplitude fluctuations is a special case or whether other, more common activities display this phenomenon. The underlying information conveyed by this inconsistency is also unknown. Therefore, the relationship between temporal amplitude fluctuations and nonlinear dynamics requires a more comprehensive investigation.

The tDES is easy to understand in terms of theoretical concepts, and is easy to implement in both software and hardware. However, its simplicity also has some drawbacks. The main parameters that can be adjusted are the delay and the threshold, which lack a certain flexibility. An increase in delay corresponds to a reduction in the sampling frequency and a decrease in the signal resolution, which generally reduces the DES value of the signal and is limited by the Nyquist sampling rate. Threshold filtering extends the application of the DES and improves its noise resistance, but reduces the accuracy with which quantitative amplitude fluctuations can be identified. Currently, we focus on the amplitude fluctuation of signals from same physiopathological states (e.g., the epileptic seizures and sleep stages). The determination of different symptoms can be made within a certain physiological activity. However, same results of quantitative amplitude fluctuation may occur for EEGs from different physiopathological conditions, so the level of refinement is not yet sufficiently high. Another limitation of our research lies in the outdated annotations at the polysomnographic database. Expert annotations on sleep stage are determined according to the outdated R\&K criteria \cite{Rechtsch1968} rather than the updated AASM manual \cite{AASM}. Furthermore, sleep stage 4 and the well-known rapid eye movement (REM) were not included due to the limitations of the public EEG recordings. Therefore, our findings should be validated using a larger and more representative amount of sleep EEG.

\section{Conclusions}
To address the original DES problem, we proposed the tDES and evaluated the quantifier using synthetic signals and real-world sleep EEGs. Our main findings are highlighted below:

Differential data reflect the instantaneous slope of time series. By measuring the distribution of differences no larger than a specified threshold, tDES improves the original DES and extends its application in quantitative amplitude fluctuations of physiological series.

Raw sleep EEG has a large number of neighbouring equal values, which increase significantly with increasing sleep depth. However, the detrend-filtered sleep EEG data has no equal values, leading to failure of the original DES.

The tDES, particularly with an adaptive threshold, reliably detects the reduced amplitude fluctuations in sleep EEGs with the progression of sleep stages, and has the advantages of insensitivity to the starting point and low data length requirements.

\section*{Acknowledgements}
The project is supported by the Natural Science Foundation of Jiangsu Province (Grant No.BK20220383), Natural Science Research of Jiangsu Higher Education Institutions of China (Grant No.22KJB110003), Natural Science Foundation of Nanjing University of Posts and Telecommunications (Grant Nos. NY221142, NY222172), Shanghai Municipal Science and Technology, China Major Project (Grant No. 2018SHZDZX01), Key Laboratory of Computational Neuroscience and Brain-Inspired Intelligence (LCNBI) and ZJLab.

\section*{References}
\bibliographystyle{iopart-num}
\bibliography{mybibfile}

\end{document}